SpringerPlus

**RESEARCH** **Open Access**

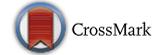

# Area/latency optimized early output asynchronous full adders and relative-timed ripple carry adders

P. Balasubramanian[1*] and S. Yamashita[2]

*Correspondence: balasubramanian@ntu.edu.sg
[1] School of Computer Engineering, Nanyang Technological University, 50 Nanyang Avenue, Singapore 639798, Singapore
Full list of author information is available at the end of the article

**Abstract**

This article presents two area/latency optimized gate level asynchronous full adder designs which correspond to early output logic. The proposed full adders are constructed using the delay-insensitive dual-rail code and adhere to the four-phase return-to-zero handshaking. For an asynchronous ripple carry adder (RCA) constructed using the proposed early output full adders, the relative-timing assumption becomes necessary and the inherent advantages of the relative-timed RCA are: (1) computation with valid inputs, i.e., forward latency is data-dependent, and (2) computation with spacer inputs involves a bare minimum constant reverse latency of just one full adder delay, thus resulting in the optimal cycle time. With respect to different 32-bit RCA implementations, and in comparison with the optimized strong-indication, weak-indication, and early output full adder designs, one of the proposed early output full adders achieves respective reductions in latency by 67.8, 12.3 and 6.1 %, while the other proposed early output full adder achieves corresponding reductions in area by 32.6, 24.6 and 6.9 %, with practically no power penalty. Further, the proposed early output full adders based asynchronous RCAs enable minimum reductions in cycle time by 83.4, 15, and 8.8 % when considering carry-propagation over the entire RCA width of 32-bits, and maximum reductions in cycle time by 97.5, 27.4, and 22.4 % for the consideration of a typical carry chain length of 4 full adder stages, when compared to the least of the cycle time estimates of various strong-indication, weak-indication, and early output asynchronous RCAs of similar size. All the asynchronous full adders and RCAs were realized using standard cells in a semi-custom design fashion based on a 32/28 nm CMOS process technology.

## Background

Asynchronous design[1] is considered to be a viable alternative and/or a necessary supplement to mainstream synchronous design by the Semiconductor Industry Association's ITRS design roadmap (ITRS 2013) due to several reliability and variability issues which have become more prominent and demand more attention in the nanoelectronics era.

---

[1] Asynchronous design, in this work, primarily refers to the design paradigm which uses a delay-insensitive (here, dual-rail) code for data representation and processing, and utilizes the 4-phase return-to-zero handshaking for communication.

Random dopant/atomistic fluctuations, sub-wavelength lithography, high heat flux, electro-migration, hot carrier effects, negative bias temperature instability, stress-induced variation, electrostatic discharge, process-induced defects, and metrology and other manufacturing defects are complicated issues (Kundu and Sreedhar 2010) which have become more pronounced in the nanoelectronics regime compared to the microelectronics regime and are indeed difficult to deal with. To overcome these issues, several material-level, process-level, device-level, circuit-level, and system-level solutions have been developed and further technological developments are forecast (ITRS 2013).

At the circuit/system level, asynchronous design methods have been drawing continued interest from the research community over the past few decades due to several inherent advantages such as low noise (Paver et al. 1998) and almost nil electro-magnetic interference (Bouesse et al. 2004), greater modularity (van Kees Berkel et al. 1999), capacity to withstand process, temperature, and parametric variations (Kulikowski et al. 2008; Chang et al. 2010), consumption of power only when and where active (van Kees Berkel et al. 1999; Akgun et al. 2010), and being self-checking (David et al. 1995). Low noise and electro-magnetic compatibility imply that asynchronous circuits are inherently resistant to side channel attacks (Yu et al. 2003; Sokolov et al. 2005) and are therefore preferable for secure environments demanded in cyber physical systems, banking and financial applications, and other consumer and industrial applications. Modularity, also known as design reusability, and the ability to tolerate process, temperature and parametric variations signifies that asynchronous circuits are well placed to cope with statistical timing analysis and reliability issues whilst delivering a good average case performance (Sparsø and Furber 2001). Due to power consumption only on demand, depending on when and where required, asynchronous circuits form a natural choice for low power VLSI designs where complimentary design strategies such as multiple supplies, multiple thresholds, and dynamic voltage and/or frequency scaling may be employed to leverage the maximum benefits from an asynchronous design.

While there exists many classes of asynchronous circuits/systems (Sparsø and Furber 2001; Myers 2001), relative-timing (Stevens et al. 2003) was proposed and suggested to be a very efficient asynchronous design style which can aggressively optimize area, delay, and power parameters much more than any other asynchronous design method. This was validated through the relative-timed design of an asynchronous instruction length decoder in (Stevens et al. 2003). However, design metric optimizations are achieved by relative-timed designs at the expense of trading off robustness, i.e., by incorporating certain timing assumptions. However, it should be noted that timing assumptions are implicit in robust asynchronous design methods such as isochronic forks (Martin 1990; Martin and Prakash 2008) in quasi-delay-insensitive designs (Toms 2006; Balasubramanian 2010), which form the weakest compromise to delay-insensitivity (van Berkel 1992), and zero or negligible wire delays in speed-independent designs (Beerel and Meng 1992; Kondratyev et al. 1994; Keller et al. 2009), while timing assumptions are made explicit in the case of relative-timing to optimize the design metrics.

This article presents two robust early output asynchronous full adder designs which when cascaded to form a ripple carry adder (RCA) gives rise to the need for incorporating relative-timing assumptions but paves the way for effective optimization of design metrics in comparison with various strong-indication, weak-indication, and early output asynchronous RCAs. Moreover, the cycle time of the proposed early output full adders based asynchronous RCAs is the least. Further, optimizations in design metrics and cycle time are achieved without sacrificing the power advantage.

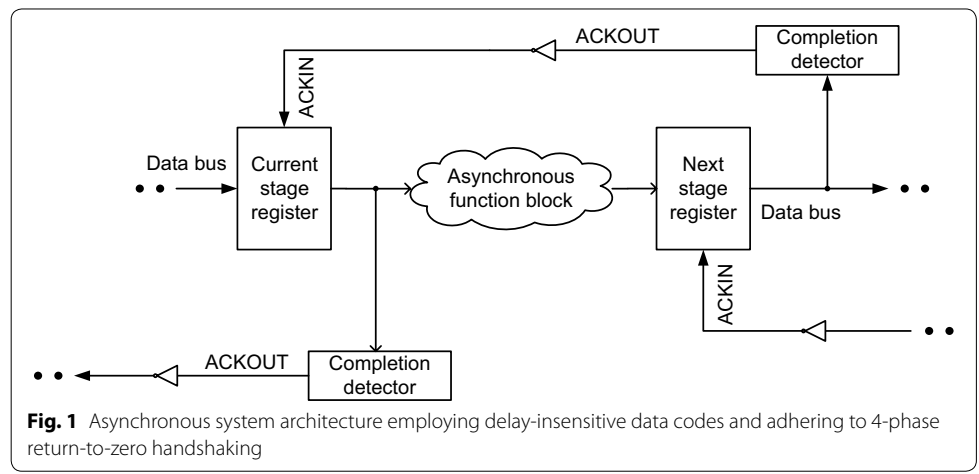

**Fig. 1** Asynchronous system architecture employing delay-insensitive data codes and adhering to 4-phase return-to-zero handshaking

The remainder of this article is organized into six sections. A short description of the standard asynchronous system architecture is given in "Asynchronous system architecture: background" section. "Asynchronous function blocks: classification" section classifies the various asynchronous function blocks and illustrates their timing characteristics on the basis of their input–output behavior. "Early output logic and relative-timing" section describes the concept of early output by translating the Seitz's weak-indication full adder into an early output design, and illustrates the relative-timing discipline based on an example 2-bit RCA constructed using the modified early output version of Seitz's weak-indication full adder. "Early output asynchronous full adders: proposed designs" section presents the proposed early output full adder designs and explains their operation. Section "Results and discussion" provides the simulation results obtained for a number of strong-indication, weak-indication, early output, and relative-timed 32-bit RCAs, which are constructed using different full adder designs including the proposed ones. Also, a theoretical estimation of cycle times for various asynchronous RCAs is provided in this section by assuming different carry-propagation lengths. "Conclusions" section finally concludes this article.

## Asynchronous system architecture: background

The standard architecture (Sparsø and Furber 2001) of a (robust) input–output mode asynchronous system[2] is shown in Fig. 1, which features a centrally located asynchronous function block. The asynchronous function block forms the heart of the asynchronous system that comprises the data path which is responsible for data processing. The asynchronous function block may be strongly indicating, weakly indicating, or early output. Given this, an asynchronous function block may fully or partly indicate (i.e., acknowledge) the arrival of the incoming primary inputs. The function block of an asynchronous system is deemed to be equivalent to the combinational logic of a synchronous system.

---

[2] The asynchronous system architecture discussed here is inherently robust. A non-robust asynchronous system, such as bundled-data system, features a separate request signal, which is not embedded within the data wires.

The asynchronous system shown in Fig. 1 employs delay-insensitive (here, dual-rail) data codes for data representation and processing, and the 4-phase return-to-zero (RTZ) protocol for handshaking (communication). The dual-rail code is the simplest member of the family of delay-insensitive *m*-of-*n* codes (Verhoeff 1988), where *m* wires are asserted 'high' (i.e., binary 1) out of a total of *n* wires to represent data. As per the dual-rail code, a data wire D is encoded into two wires viz. D0 and D1, where D = 1 is represented by D1 = 1 and D0 = 0, and D = 0 is represented by D0 = 1 and D1 = 0. When D1 and D0 signify a binary value of 0 or 1, according to the assignments mentioned, it is called 'valid data'. The state of both D0 and D1 being equal to 0 is called the 'spacer'. Note that both D0 and D1 cannot simultaneously transition to 1 as it is illegal and invalid since the coding scheme adopted is unordered (Bose 1991), where no codeword is allowed to be a subset of another codeword.

Referring to Fig. 1, the 4-phase RTZ handshake protocol is explained as follows.[3] The dual-rail data bus that feeds the current stage register is initially in the spacer state, and the common acknowledge input (ACKIN) for the current stage register is binary 1, since the common acknowledge output (ACKOUT) provided by the next stage register is binary 0. The current stage register now transmits a codeword (i.e., valid data). This results in low to high transitions on the bus wires (i.e., any one of the rails of all the dual-rail signals is asserted as binary 1) which feed the function block. After the next stage register receives a codeword subsequent to data processing in the function block, it drives the ACKOUT wire to binary 1, and the ACKIN wire assumes binary 0. The current stage register waits for the ACKIN signal to become 0 and then resets the data bus, i.e., the data bus feeding the function block is driven to the spacer state. After an unbounded but finite and positive amount of time taken for the resetting of the function block and the passage of the spacer to the following register stage, the next stage register drives the ACKOUT (ACKIN) to 0 (1). With this, a data transaction is said to be complete, and the system is ready to commence the next data transaction. The application of input data in the asynchronous system follows the sequence: valid data–spacer–valid data–spacer, and so forth.

## Asynchronous function blocks: classification

Asynchronous function blocks, constructed using delay-insensitive codes, can be classified as strongly indicating, weakly indicating, and early output, in order, according to their robustness hierarchy and depending upon the nature of their indication (i.e., acknowledgment). The differences between the properties of strong-indication, weak-indication, and early output asynchronous function blocks shall be explained using an illustrative timing diagram shown as Fig. 2.

### Strongly indicating asynchronous function block

A strongly indicating asynchronous function block waits for all the valid/spacer primary inputs to arrive and then starts to produce the requisite valid/spacer primary outputs (Seitz 1979; Balasubramanian and Edwards 2008a). The primary input–output sequencing constraints corresponding to strong-indication are:

- All the primary inputs attain valid/spacer state before any primary output attains valid/spacer state respectively
- All the primary outputs should have attained valid/spacer state before any primary input attains spacer/valid state respectively

---

[3] This explanation remains valid for any delay-insensitive data encoding scheme.

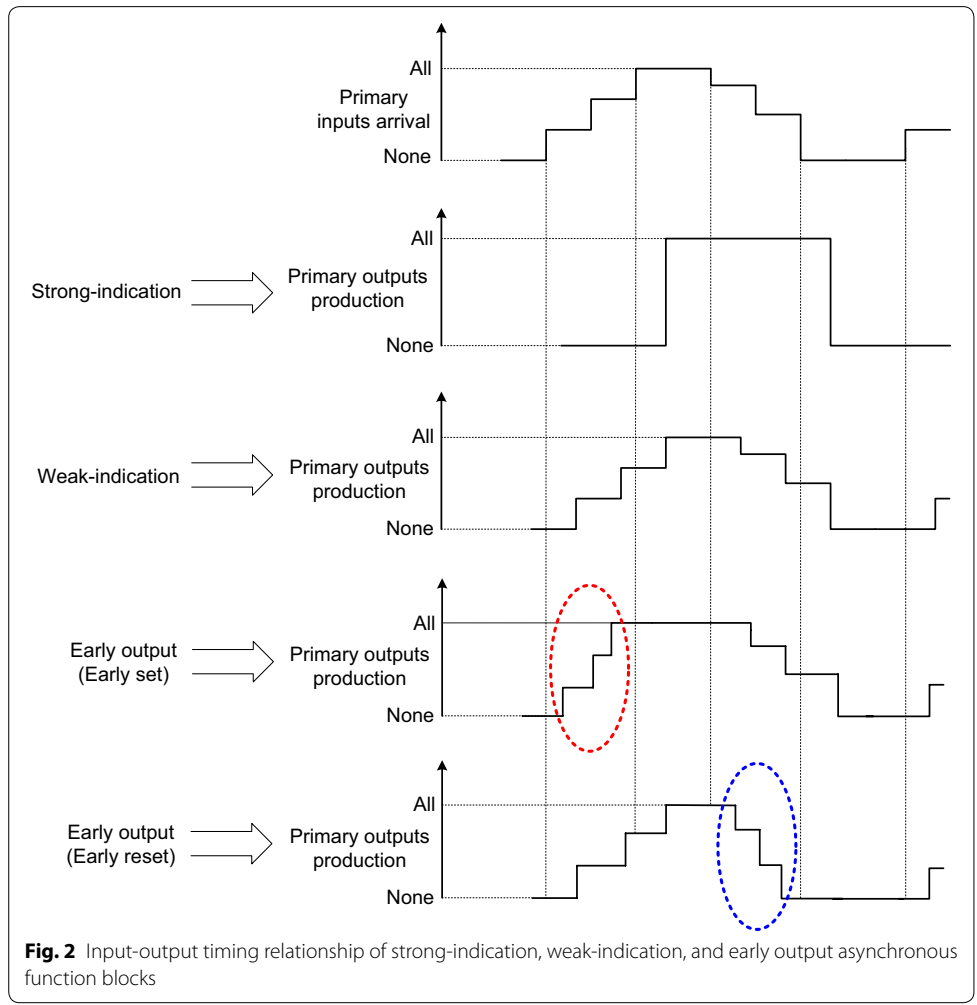

**Fig. 2** Input-output timing relationship of strong-indication, weak-indication, and early output asynchronous function blocks

**Weakly indicating asynchronous function block**

A weakly indicating asynchronous function block tends to produce valid/spacer primary outputs subsequent to the arrival of even a subset of the valid/spacer primary inputs (Seitz 1979; Balasubramanian and Edwards 2008b). However, the production of at least one valid/spacer primary output is withheld until all the valid/spacer primary inputs have arrived. The primary input–output sequencing constraints in this case are:

- Some valid/spacer primary outputs are produced after some valid/spacer primary inputs arrive respectively

- All the valid/spacer primary inputs should have arrived before all the corresponding valid/spacer primary outputs are produced
- All the valid/spacer primary outputs should have been produced before any subsequent spacer/valid primary inputs arrive respectively

**Early output asynchronous function block**

The early output asynchronous function block (Brej and Garside 2003; Balasubramanian 2011) is relaxed compared to the strong and weak-indication function block counterparts since all the valid/spacer primary outputs may be produced in this case subsequent to the arrival of just a subset of the corresponding valid/spacer primary inputs. The primary input–output sequencing constraints pertaining to early output logic are:

- All the valid/spacer primary outputs may be produced with the arrival of just a subset of the corresponding valid/spacer primary inputs
- After all the valid/spacer primary inputs arrive, the primary outputs continue to maintain the same valid/spacer state respectively

Figure 2 illustrates two kinds of early output asynchronous function blocks which permit either early reset or early set. An early set type asynchronous function block would produce all the valid primary outputs subsequent to the arrival of even a subset of the valid primary inputs without waiting for the arrival of all the valid primary inputs. The early set nature of an asynchronous function block is highlighted by the red oval in Fig. 2. On the other hand, the early reset asynchronous function block could drive all the primary outputs to spacers following the assumption of the spacer state by even a subset of the primary inputs without waiting for all the primary inputs to become spacers. The early reset nature of an asynchronous function block is highlighted by the blue oval in Fig. 2.

Figure 3 illustrates the concept of early output through simple logic gates with Z and L being designated as the primary outputs, and X, Y, J, and K being labeled as the primary inputs. The low to high up going transition (binary 0 to 1) on a wire M is represented by M↑, and the down going (binary 1 to 0) transition on the wire M is denoted by M↓. Referring to the AND gate shown in Fig. 3, it can be seen that X↓ is followed by Z↓ despite Y maintaining the steady-state of 1, which is indicative of an early reset. Referring to the OR gate shown in Fig. 3, it can be seen that J↑ is followed by L↑ despite K maintaining 0 as the steady-state, which is reflective of an early set.

**Early output logic and relative-timing**

To explain the concept of early output with respect to an asynchronous function block, let us consider the Seitz's weak-indication full adder (Seitz 1979), shown in Fig. 4.

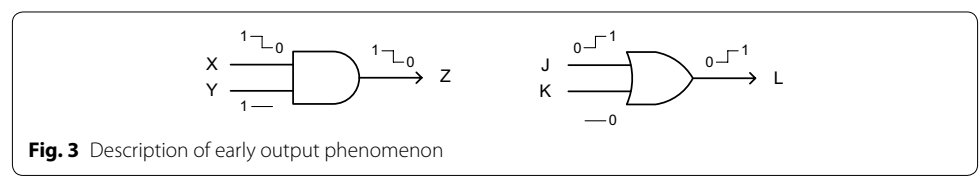

**Fig. 3** Description of early output phenomenon

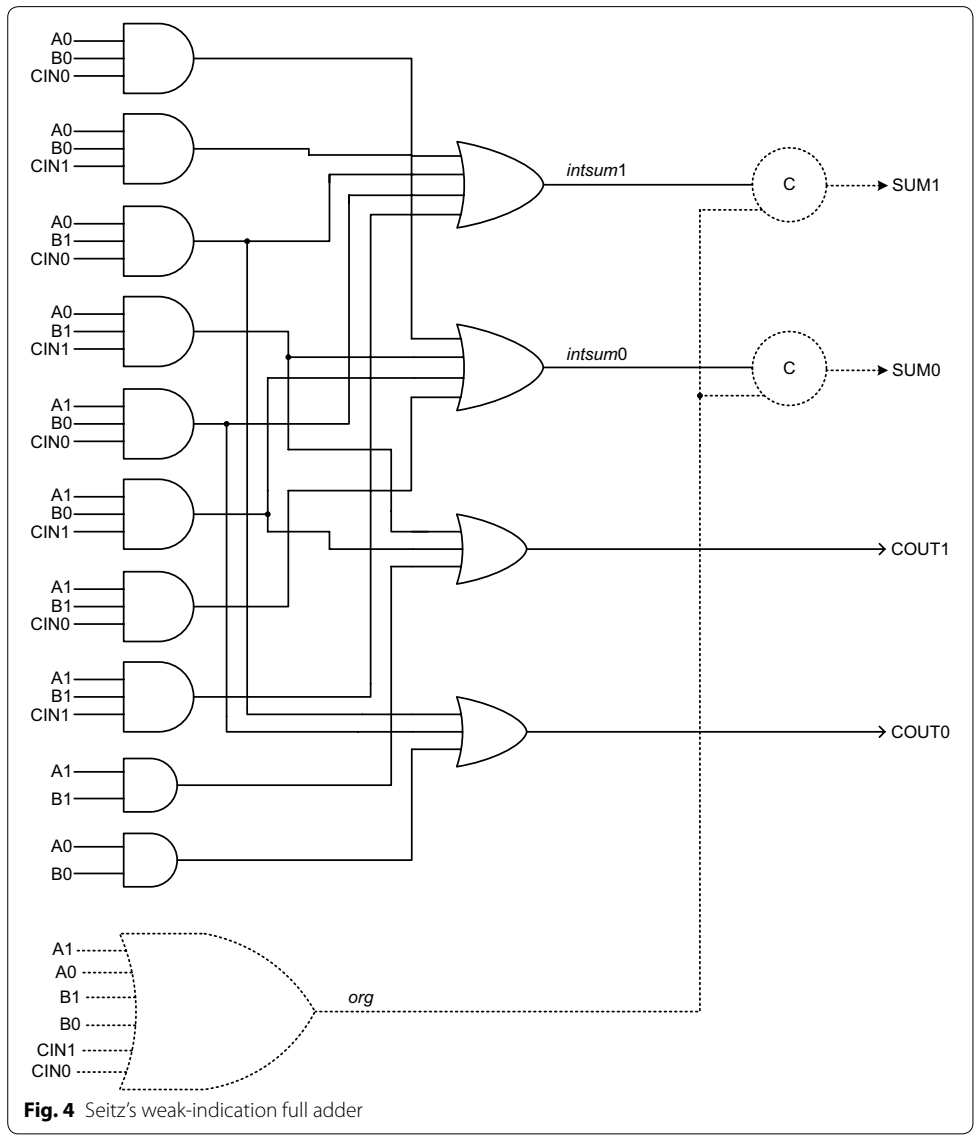

**Fig. 4** Seitz's weak-indication full adder

In Fig. 4, A1, A0, B1, B0, CIN1 and CIN0 represent the dual-rail primary inputs, while SUM1, SUM0, COUT1 and COUT0 denote the dual-rail primary outputs. The logic equations governing the dual-rail encoded full adder outputs are given by (1) to (4). It may be noted that the product terms constituting the below expressions are mutually disjoint (Balasubramanian and Edwards 2010; Balasubramanian et al. 2012a; Balasubramanian and Mastorakis 2010), i.e., the logical conjunction of any pair-wise products equals null.

$$\text{SUM1} = \text{A0B0CIN1} + \text{A0B1CIN0} + \text{A1B0CIN0} + \text{A1B1CIN1} \tag{1}$$

$$\text{SUM0} = \text{A0B0CIN0} + \text{A0B1CIN1} + \text{A1B0CIN1} + \text{A1B1CIN0} \tag{2}$$

$$\text{COUT1} = \text{A0B1CIN1} + \text{A1B0CIN1} + \text{A1B1} \tag{3}$$

$$COUT0 = A0B1CIN0 + A1B0CIN0 + A0B0 \qquad (4)$$

From (1) to (4), it can be seen that the sum outputs depend upon all the primary inputs while the carry outputs may not. For example, when the carry-propagate condition occurs, either A0 and B1 or A1 and B0 would transition to binary 1, and COUT1 or COUT0 would have to wait for the arrival of the incoming carry viz. CIN1 or CIN0. However, when carry-generation occurs (i.e., A1 = B1 = 1), COUT1 can assume 1 without having to wait for the incoming carry. Similarly, when carry-kill occurs (i.e., A0 = B0 = 1), COUT0 can become 1 regardless of the incoming carry. Since AND gates are used to realize the product terms in Fig. 4, when spacer data are applied, even with a single dual-rail input pair say A1 and A0 assuming binary 0, the outputs of the AND gates present in the first logic level of Fig. 4, and the outputs of the OR gates present in the second logic level of Fig. 4 would become 0. Hence, the internal outputs *intsum*1 and *intsum*0, which are logically equivalent to SUM1 and SUM0 respectively, would become 0, and so would the primary carry outputs COUT1 and COUT0. However, *intsum*1 and *intsum*0 are joined to another internal output viz. *org* of the 6-input OR gate through two 2-input C-elements[4] as shown in Fig. 4, where the C-elements are represented by the circles with the marking 'C' on their periphery.

Since the internal output *org* would become 0 only after all the primary inputs have become spacers, and because the C-elements would output 0 on both SUM1 and SUM0 only after *org*, *intsum*1 and *intsum*0 have all become 0s, the dual-rail sum output (SUM1/SUM0) would have to wait for the arrival of all the spacer inputs before the production of spacers through them which is not the case with the dual-rail carry output viz. COUT1/COUT0. The 6-input OR gate shown in dotted lines in Fig. 4 would confirm the complete arrival of all the spacer inputs. When valid data are applied and presuming only A1 has transitioned to 1, *org* might attain 1 without waiting for the assertion of the remaining dual-rail primary inputs. Nevertheless, at this juncture, the arrival of the requisite primary inputs is necessary for a valid output production through the AND gate(s) present in the first logic level, and eventually the sum output would confirm the complete arrival of all the requisite primary inputs.

If the 6-input OR gate and the two 2-input C-elements shown in dotted lines are removed from Fig. 4, the internal outputs *intsum*1 and *intsum*0 would no more be present and would simply be referred to as the primary sum outputs viz. SUM1 and SUM0 as shown in Fig. 5. Given this, when valid data are applied, all the requisite valid inputs have to arrive before a valid output can be produced on SUM1 or SUM0. This is not necessarily the case with the carry outputs because the carry output logic may take advantage of the carry-generate/carry-kill condition as discussed earlier. However when spacer data are applied, even with a single dual-rail input pair assuming the spacer state, all the primary outputs viz. SUM1, SUM0, COUT1 and COUT0 could assume the spacer state. Hence, even with a subset of the primary inputs becoming spacers, the entire full adder shown in Fig. 5 could be reset unlike the full adder shown in Fig. 4, which requires all the primary inputs to become spacers. Thus with the stated modifications, i.e., after removal of the 6-input OR gate and the two 2-input C-elements shown in

---

[4] The C-element outputs a 1 (0) only if all its inputs are 1 (0). It retains the existing steady-state if the applied inputs are different.

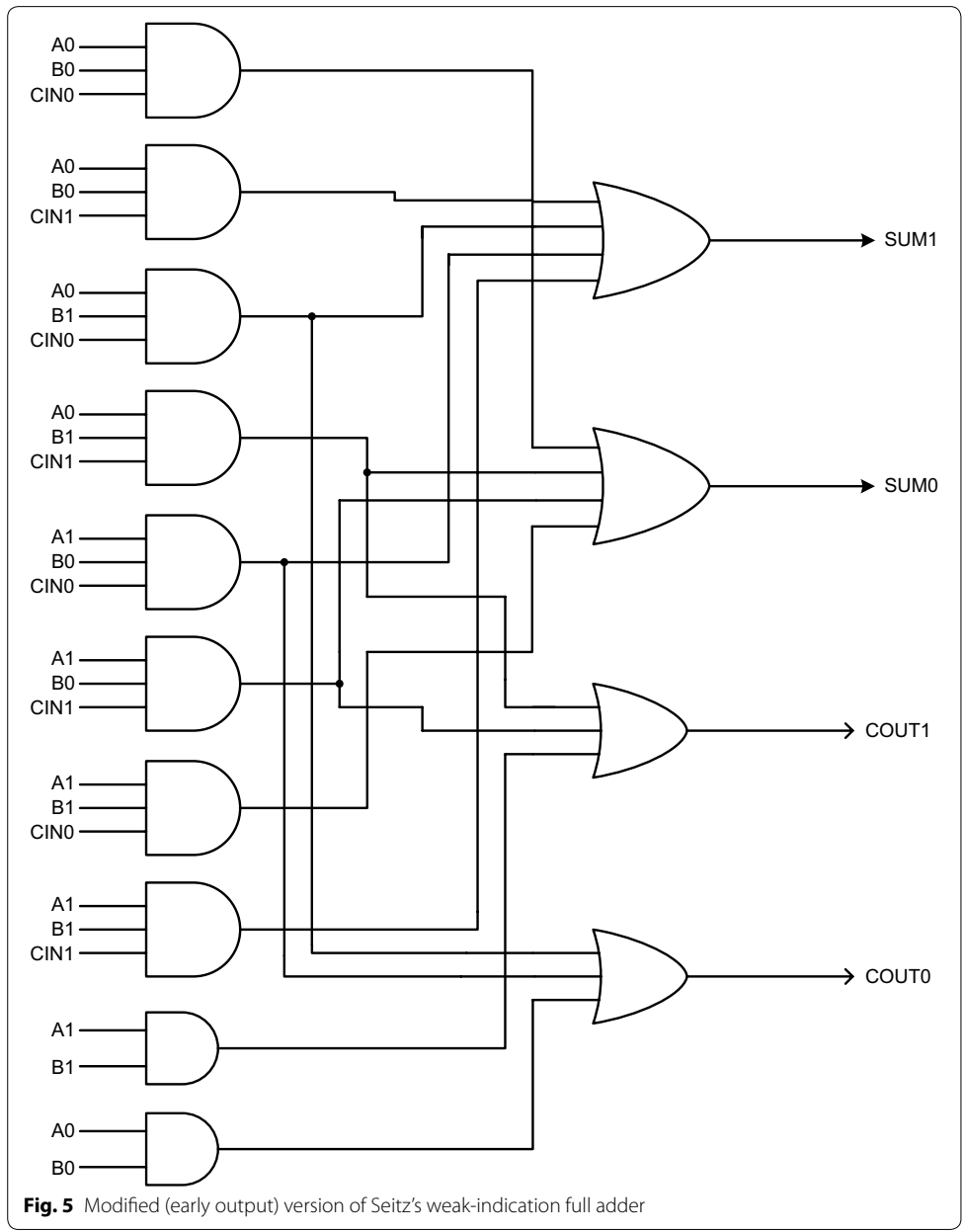

**Fig. 5** Modified (early output) version of Seitz's weak-indication full adder

Fig. 4, the Seitz's weak-indication full adder would be transformed into an early output i.e., early reset type design as shown in Fig. 5. If the full adder shown in Fig. 5 is used to replace the asynchronous function block shown in Fig. 1, the indication of all the primary inputs regardless of whether they assume valid or spacer data does not become an issue since the completion detector would take care of indicating the arrival of all the primary inputs.

Figure 6 shows an example 2-bit asynchronous RCA as embedded into the asynchronous system. In Figs. 6 and 7, (A11, A10), (A01, A00), (B11, B10), (B01, B00) and (C01, C00) represent the dual-rail primary inputs, while (SUM11, SUM10), (SUM01, SUM00) and (C21, C20) signify the dual-rail primary outputs. Here, (C01, C00) denotes the incoming carry to the RCA, (C11, C10) represents the internal dual-rail carry, and (C21, C20) signifies the carry overflow/carry output of the RCA.

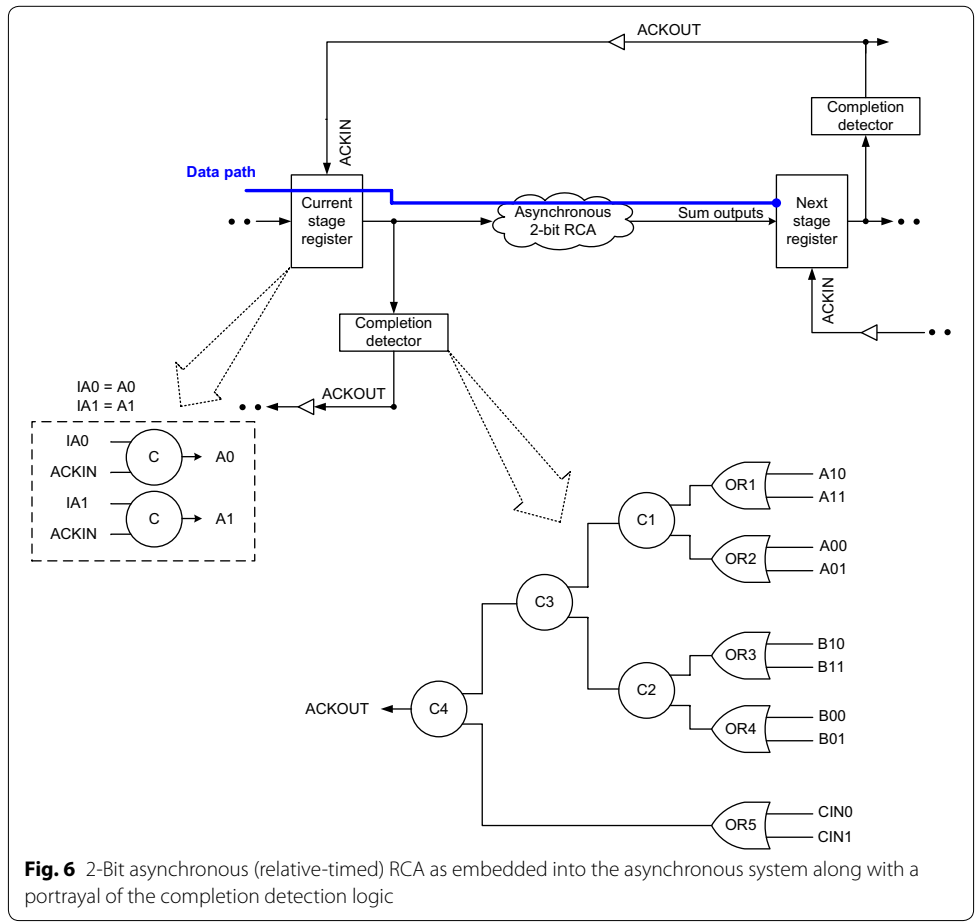

**Fig. 6** 2-Bit asynchronous (relative-timed) RCA as embedded into the asynchronous system along with a portrayal of the completion detection logic

The completion detector (Sparsø and Furber 2001), shown in Fig. 6, consists of a series of 2-input OR gates labelled as OR1 to OR5, with each 2-input OR gate dedicated to combine the respective dual-rails of an encoded input, and the outputs of such 2-input OR gates are synchronized through a C-element tree characterized by C1, C2, C3 and C4 in Fig. 6. When valid input data are supplied to the asynchronous RCA shown in Fig. 6, they are supplied to the completion detector as well. With isochronic fork assumptions imposed on all the primary input wires, the problem of wire orphan viz. unacknowledged signal transition on a wire gets eliminated. Note that a fork is called isochronic (Martin 1990, 2009) only if all the branches of the fork experience similar signal transitions occurring concurrently. An isochronic fork forms the weakest compromise to delay-insensitivity (Martin 1990). Upon application of valid input data, at least one of the dual-rails of all the primary inputs to the 2-bit RCA would transition to 1, which implies all the 2-input OR gates viz. OR1 to OR5 in Fig. 6 will transition to 1. Only after these transitioning happen, all the 2-input C-elements viz. C1 to C4 in Fig. 6 will output 1, thus acknowledging the arrival of valid data on all the primary inputs to the RCA through the production of 1 on ACKOUT. The same phenomenon repeats in the subsequent RTZ phase wherein only after all the primary inputs to the RCA shown in

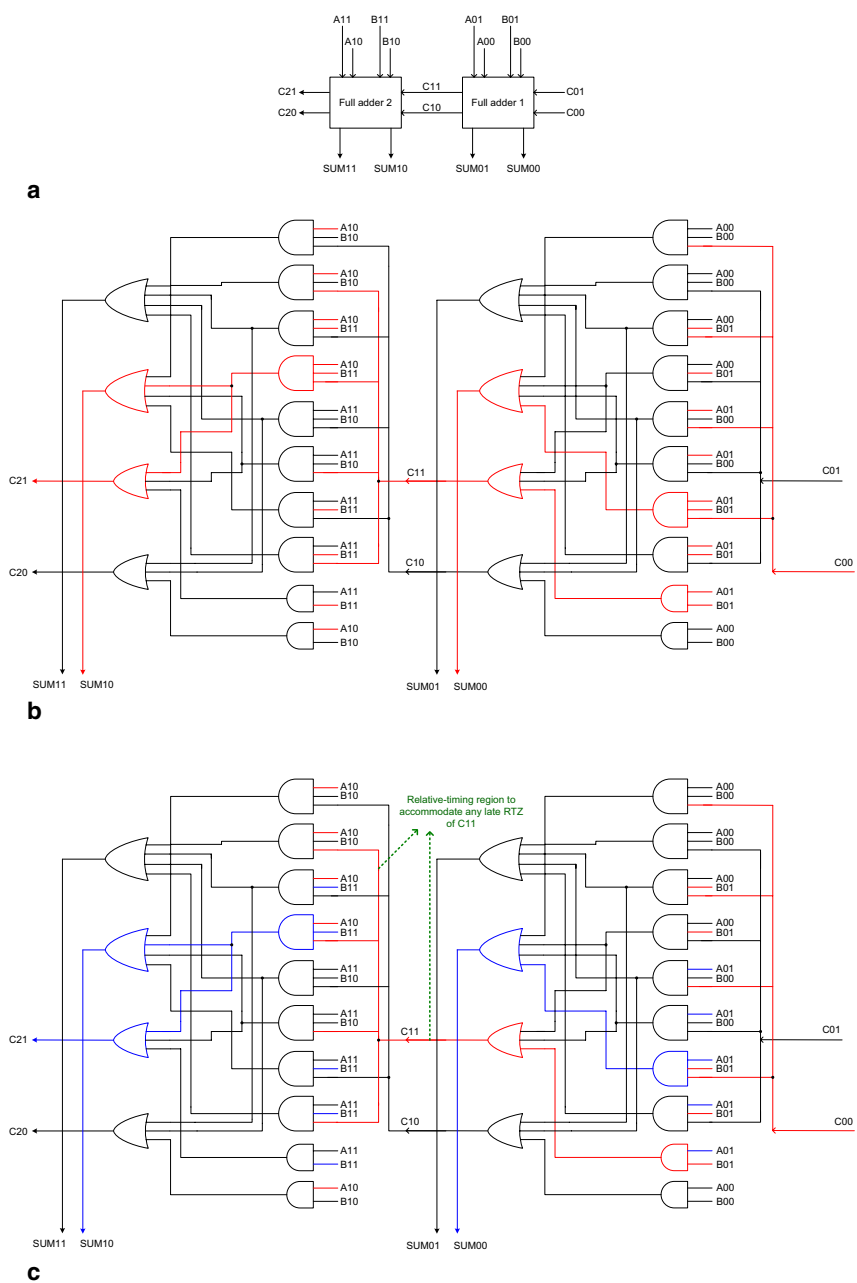

**Fig. 7** 2-bit relative-timed RCA composed using early output full adders such as the one shown in Fig. 5. The relative-timing assumption made is that C11↓ or C10↓ precedes SUM11↓ or SUM10↓ during RTZ. **a** 2-bit relative-timed RCA. The internal details and signal transitions corresponding to application of valid and spacer data is shown in **b**, **c** respectively. **b** Example application of valid input data in the 2-bit relative-timed RCA constructed by cascading two stages of the early output version of Seitz's full adder. Carry generation in least significant stage: A01 = B01 = 1 with C00 = 1 assumed; SUM00 = 1 and C11 = 1 results. Carry propagation in most significant stage: A10 = B11 = 1 with C11 = 1 from the previous stage; SUM10 = 1 and C21 = 1 results. **c** The following RTZ phase in the 2-bit relative-timed RCA with only a partial reset of the primary inputs. Early RTZ of A01 and B11 alone could facilitate the simultaneous RTZ of SUM00, SUM10 and C21, and the late RTZ of A10 and B01 would be indicated by the completion detector. RTZ of the internal carry C11 may not be acknowledged! Hence, relative-timing assumption for the regions pointed to in *green* is necessary to avoid the potential problem of orphans

Fig. 6 RTZ, will the outputs of OR gates OR1 to OR5 RTZ, followed by the RTZ of the outputs of C-elements viz. C1 to C4, thereby ACKOUT also returns to 0 indicating that all the primary inputs to the RCA have returned to 0. Thus, while the asynchronous RCA shown in Fig. 6 may RTZ eagerly, the completion detector preceding it would promptly indicate the arrival of both valid and spacer data on the primary inputs to the RCA in the corresponding phases.

Although the completion detector shown in Fig. 6 would take care of indication of all the primary inputs supplied to the RCA, an issue would arise when the asynchronous 2-bit RCA shown in Fig. 6 is considered to be relative-timed. For the present discussion, the full adders constituting the 2-bit RCA in Fig. 6 are considered to be the one shown in Fig. 5 viz. the early output version of Seitz's weak-indication full adder. The important issue that has to be addressed is the indication of the internal carry signal which might tend to propagate from one full adder stage to another if the carry-propagate condition becomes active in the RCA, which gives rise to the need for the assumption of relative-timing. To discuss this issue, we refer to Fig. 7.

Figure 7a shows the block schematic of two early output full adders cascaded to form a 2-bit relative-timed RCA, as shown in Fig. 6. The two early output full adders are considered to be similar to the one shown in Fig. 5. Figure 7b, c show the internal circuit details of the 2-bit asynchronous (relative-timed) RCA constructed using the early output version of Seitz's weak-indication full adder shown in Fig. 5. The full adder on the right-hand-side of Fig. 7b, c represents the least significant stage, and the full adder on the left-hand-side of Fig. 7b, c represents the most significant stage.

The red lines shown in Fig. 7b signify an example application of the valid input data: $A0_1 = B0_1 = 1$, and $A1_0 = B1_1 = 1$ with carry input of $C0_0 = 1$ assumed, and the corresponding signal transitions on the inputs and internal nodes leading to the production of the required sum and carry outputs viz. $SUM1_0 = SUM0_0 = 1$ and $C2_1 = 1$, which are also highlighted in red. Figure 7c is a replica of Fig. 7b, where the RTZ of a subset of the primary inputs, intermediate outputs, and the primary sum and carry outputs is highlighted by the blue lines. Figure 7c is used to illustrate how even a partial RTZ of the primary inputs viz. $B1_1$ and $A0_1$ returning to 0 could result in the RTZ of $SUM1_0$, $SUM0_0$ and $C2_1$, which were 1 earlier, portraying the early reset nature of the relative-timed RCA. In this case, the late RTZ of $A1_0$ and $B0_1$ would be acknowledged by the completion detector preceding the 2-bit relative-timed RCA as shown in Fig. 6. Thus the RTZ of all the primary inputs excepting the internal carry output ($C1_1$/$C1_0$) would be acknowledged by the RCA and the completion detector, but the RTZ of the internal carry $C1_1$ may not be acknowledged. The circuit portion pointed to in green in Fig. 7c specifies the region where the relative-timing assumption has to be imposed in order to avoid any ambiguity about the RTZ of the internal carry output viz. $C1_1$ thus potentially preventing any gate/wire orphan occurrence. Orphans are basically unacknowledged signal transitions which may occur on gate output(s) or wire(s) and if left unattended might affect the robustness of an asynchronous circuit/system. To overcome the orphan problem, timing assumptions are usually necessary and these may be implicit or explicit in the design. For a clear explanation of gate and wire orphans, the reader is referred to (Balasubramanian 2016).

Thus with reference to Fig. 7, in general, for application of valid input data, the RCA sum and carry outputs would be produced subsequent to the arrival of the primary RCA inputs including the incoming carry. The sum outputs would indicate the arrival of all the valid primary inputs while the carry outputs may or may not depending upon whether carry-propagate or carry-generate or carry-kill condition is activated. The completion detector preceding the 2-bit RCA, shown in Fig. 6, may provide multiple acknowledgments for the valid primary inputs supplied to the RCA. When spacer data is applied in the RTZ phase, following the RTZ of even one primary input corresponding to both the full adders in the RCA, the primary sum and carry outputs could RTZ independent of each other. Even if the remainder of the spacer inputs arrives lately, their arrival would be duly acknowledged by the completion detector preceding the asynchronous RCA, as shown in Fig. 6. Further, the sum outputs of both the full adders in the RCA could RTZ in parallel, while the internal carry reset and its propagation may or may not happen simultaneously. Moreover, the late RTZ of the internal dual-rail carry (C11/C10) may also not be acknowledged. Hence, to avoid any ambiguity with respect to the RTZ of the internal carry and to avoid the potential problem of gate orphan, a relative-timing assumption is imposed for the RTZ phase that C11↓ or C10↓ is presumed to have happened before SUM11↓ or SUM10↓ happens. Note that the RTZ of the internal carry would also result in a similar RTZ of the primary sum output, i.e., the successive transitions would be monotonically decreasing (Cortadella et al. 2004). Hence the relative-timing assumption imposed does not contradict the monotonicity of the transitions.

With respect to an $n$-bit relative-timed RCA realized using early output asynchronous full adders, an example of which was shown in Fig. 5, the essential relative-timing assumption required can be generalized as thus: the sum output of any early output full adder stage is presumed to have attained the spacer state during a RTZ phase only after its carry input from the preceding full adder stage has become the spacer. This generalization of the relative-timing assumption concerns a maximum of only two full adder stages in an $n$-bit relative-timed RCA and that too only with respect to the RTZ phase. Also, such a relative-timing assumption may be quite conveniently realized by utilizing bigger sized gates exclusively for carry output production in the full adders or through delay-balancing so that the propagation delay encountered in directly resetting the sum output of any $(q + 1)$th full adder stage in a relative-timed RCA is equal to the propagation delay incurred in indirectly resetting the sum output of any $(q + 1)$th full adder stage through the resetting of its carry input supplied from the $q$th full adder stage.

The time taken (i.e., propagation delay) to process valid data is called forward latency, and the time taken to process spacer data is called reverse latency. The sum of forward and reverse latencies gives the cycle time. The cycle time specifies the time taken to complete a single data transaction in a 4-phase handshaking protocol. It may be noted (Balasubramanian and Mastorakis 2012) that a strong-indication RCA would encounter worst-case latency for computation of valid and spacer data; a weak-indication RCA might encounter data-dependent latency for both valid and spacer data, or data-dependent latency for valid data and a constant-time latency of 2 full adder delays for spacer data (Balasubramanian 2015) depending upon whether the weak-indication RCA comprises basic or distributed/biased weak-indication full adders; an early output RCA would encounter data-dependent latency for valid data and constant-time latency of 2 full adder delays for the spacer; and a relative-timed RCA would encounter data-dependent latency for valid data and the least constant-time latency of just 1 full adder delay for the spacer as shown in Table 1. The reverse latency of the relative-timed RCA is theoretically the least because all the constituent full adders of this RCA could be reset i.e., RTZ in parallel upon the assumption of spacer state by the corresponding primary inputs regardless of the internal carries having returned to 0 or not during the RTZ phase.

Considering an *n*-bit weak-indication RCA constructed using a cascade of distributed or biased weak-indication full adders, valid data processing would be data-dependent, while spacer data processing would entail a constant time of 2 full adder delays—one full adder delay for the RTZ of the carry output, which serves as the carry input for the successive full adder stage and triggers the RTZ of its sum output which involves another full adder delay. Therefore, the least cycle time of an *n*-bit weak-indication RCA featuring distributed or biased weak-indication full adders would be specified by the sum of forward and reverse latencies as $(m + 2) \times T_{\text{full adder}}$, where *m* represents the maximum number of full adder stages in the *n*-bit RCA through which the carry propagates, and $T_{\text{full adder}}$ denotes the propagation delay of the full adder. On the other hand, for the *n*-bit relative-timed RCA constructed using early output asynchronous full adders, its forward and reverse latencies would be specified by $m \times T_{\text{full adder}}$ and $T_{\text{full adder}}$ respectively, and thus its cycle time would be governed by an optimal value of $(m + 1) \times T_{\text{full adder}}$.

### Early output asynchronous full adders: proposed designs

The two early output asynchronous full adders proposed in this work are portrayed by Figs. 8 and 9. For ease of referencing, these two full adders shall henceforth be referred by the acronyms AOPT_EO_FA (i.e., area optimized early output full adder) and LOPT_EO_FA (i.e., latency optimized early output full adder). The former requires less number of logic gates than the latter and hence occupies less area for physical realization, while the latter enables reduced latency compared to the former on account of its physical composition. It may be noted that logic redundancy is implicit (Balasubramanian et al. 2012b) in AOPT_EO_FA while there is no redundant logic in LOPT_EO_FA. In Figs. 8 and 9, A1, A0, B1, B0, CIN1 and CIN0 represent the dual-rail primary inputs, while SUM1, SUM0, COUT1 and COUT0 denote the dual-rail primary outputs. Let us first consider the AOPT_EO_FA followed by the LOPT_EO_FA.

**Table 1 Latencies and cycle time magnitudes of strong, weak, early output, and relative-timed *n*-bit RCAs, with *m* representing the maximum length of the carry chain exercised ($m \leq n$)**

| RCA realization style | Forward latency | Reverse latency | Cycle time |
|---|---|---|---|
| Strong-indication | $n \times T_{\text{full adder}}$ | $n \times T_{\text{full adder}}$ | $2n \times T_{\text{full adder}}$ |
| Weak-indication (basic) | $m \times T_{\text{full adder}}$ | $m \times T_{\text{full adder}}$ | $2m \times T_{\text{full adder}}$ |
| Weak-indication (distributed/biased) | $m \times T_{\text{full adder}}$ | $2T_{\text{full adder}}$ | $(m + 2) \times T_{\text{full adder}}$ |
| Early output | $m \times T_{\text{full adder}}$ | $2T_{\text{full adder}}$ | $(m + 2) \times T_{\text{full adder}}$ |
| Relative-timed (proposed) | $m \times T_{\text{full adder}}$ | $T_{\text{full adder}}$ | $(m + 1) \times T_{\text{full adder}}$ |

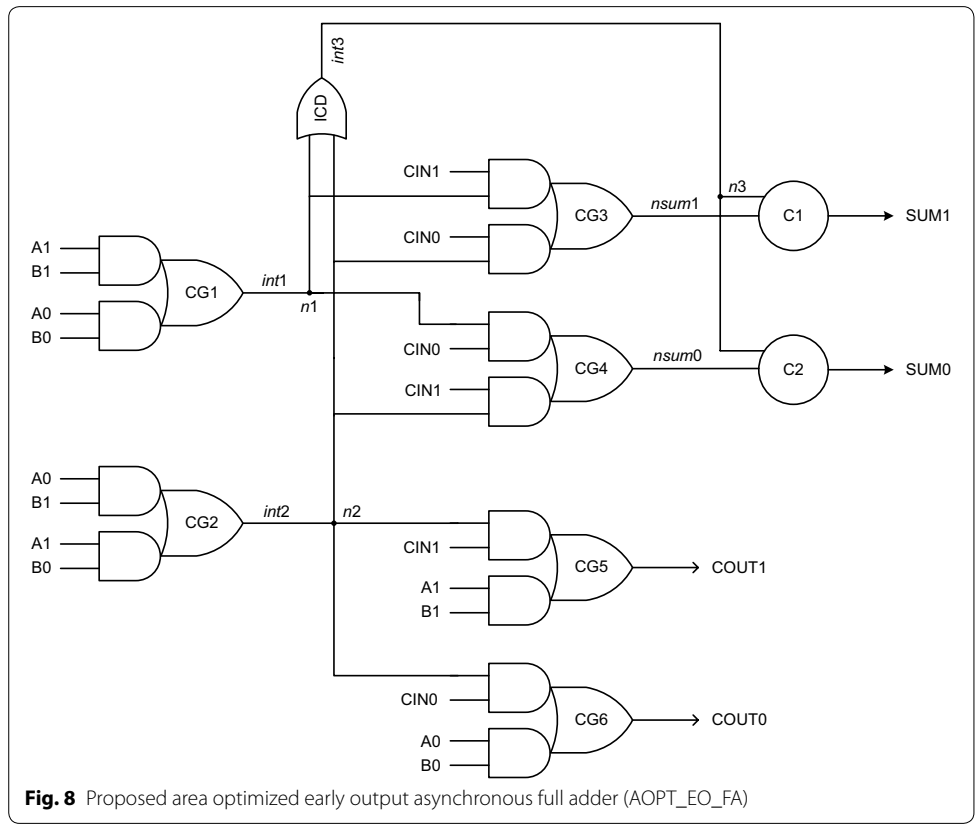

**Fig. 8** Proposed area optimized early output asynchronous full adder (AOPT_EO_FA)

The AOPT_EO_FA contains 6 AO22 cells, marked as CG1 to CG6 in Fig. 8, two 2-input C-elements marked as C1 and C2, and a 2-input OR gate marked as ICD. Of these, the AO22 gates and the C-elements are complex gates. An AO22 gate with inputs A, B, C, D and output Y implements the Boolean function Y = AB + CD. The 2-input Muller C-element, realized using the AO222 gate with feedback, implements the logic function Z = XY + (X + Y) Z, where X and Y represent the inputs and Z represents the output. There are five internal outputs in Fig. 8 viz. *int*1, *int*2, *int*3, *nsum*1 and *nsum*0, with *nsum*1 and *nsum*0 being logically equivalent to SUM1 and SUM0 respectively. The internal nodes *n*1, *n*2, and *n*3 represent isochronic fork junctions. The OR gate bearing the label 'ICD' is an internal completion detector which acknowledges a rising transition occurring on *int*1/*int*2 during the valid data phase, and the assumption of the spacer state by *int*1 and *int*2 during the RTZ phase. Both the AOPT_EO_FA and LOPT_EO_FA synthesize (1) to (4) given earlier but in structurally different forms.

Let us now consider some example input scenarios viz. carry propagation, carry generation, and carry-kill to describe the operation of AOPT_EO_FA.

- *Carry-propagate condition*: A0 = B1 = 1 (or) A1 = B0 = 1

  For either of the stated input combinations, the complex gate CG2 will be enabled and the internal output *int*2 will become 1. The internal completion detector (ICD) will also become enabled and *int*3 will become 1. Assuming CIN1 as 1, the intermediate output *nsum*0 would become 1. The C-element, marked as C2, would wait for the arrival of rising transitions on *nsum*0 and *int*3 and after their arrival would pro-duce 1 on SUM0. With *int*2 and CIN1 being 1, the carry output COUT1 would also become 1. For the inputs combination assumed, both the sum and carry outputs are found to be dependent upon the arrival of all the primary inputs. In a subsequent RTZ phase, even with A0 and/or B1 or A1 and/or B0 returning to 0, COUT1 could assume the spacer state irrespective of CIN1 assuming the spacer. Also, SUM0 could RTZ regardless of CIN1 assuming the spacer. These demonstrate the early reset nature of AOPT_EO_FA.

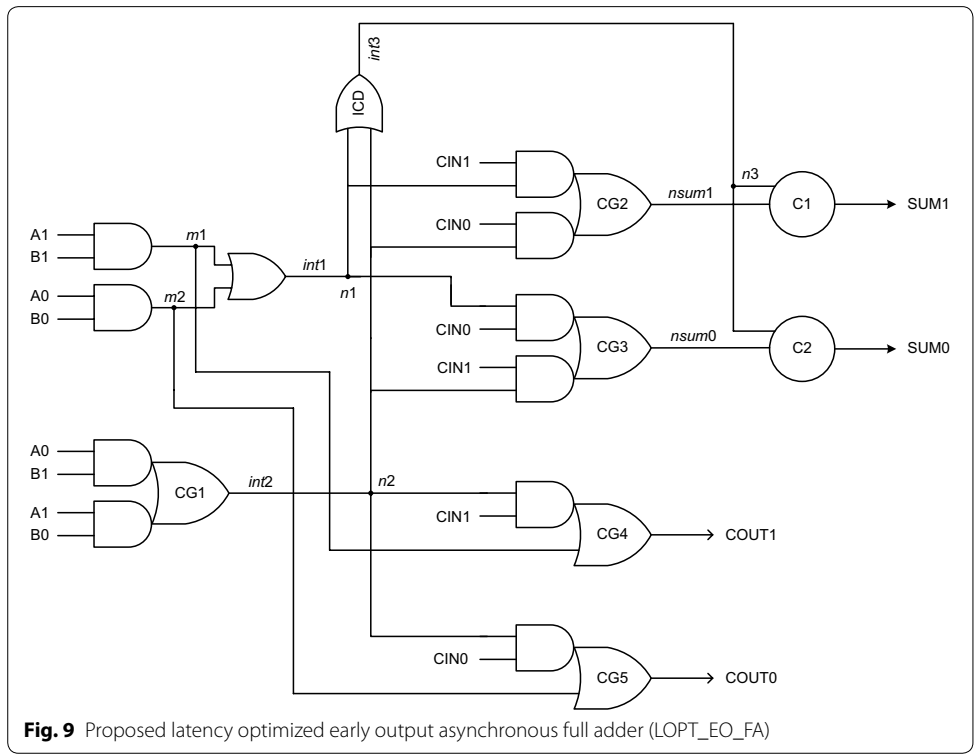

**Fig. 9** Proposed latency optimized early output asynchronous full adder (LOPT_EO_FA)

- *Carry-generate condition*: A1 = B1 = 1

  In this case, the complex gate CG5 will be enabled since A1 and B1 are 1, which will result in the generation of the carry output COUT1 = 1 regardless of whether CIN1 or CIN0 is 1. Further, the complex gate CG1 will also be enabled and a rising transition on *int*1 would be followed by a similar rising transition on *int*3. Depending upon whether CIN0 or CIN1 is asserted as 1, *nsum*0 or *nsum*1 would experience a rising transition, which when coupled with the rising transition on *n*3, results in either SUM0 or SUM1 to be asserted high. During the subsequent RTZ phase, even if A1 and/or B1 returns to 0, COUT1 and SUM1/SUM0 which were 1 earlier could RTZ irrespective of the RTZ of the carry input viz. CIN1/CIN0. These demonstrate the early reset nature of AOPT_EO_FA.

- *Carry-kill condition*: A0 = B0 = 1

  In this case, the complex gate CG6 will be enabled since A0 and B0 are 1, which will eventually result in COUT0 = 1 regardless of whether CIN1 or CIN0 is 1. Further, the complex gate CG1 will be enabled and a rising transition on *int*1 would be followed by a similar rising transition on *int*3. Depending upon whether CIN0 or CIN1 is asserted as 1, *nsum*0 or *nsum*1 would experience a rising transition, which when coupled with the rising transition on *n*3 results in either SUM0 or SUM1 to be asserted high. In the following RTZ phase, even if A0 and/or B0 returns to 0, COUT0 would RTZ and SUM0/SUM1 whichever was 1 earlier would also RTZ. Both these could happen regardless of CIN1/CIN0 having returned to 0, which again demonstrates the early reset nature of AOPT_EO_FA.

The LOPT_EO_FA is shown in Fig. 9 and is designed using 7 complex gates (3 AO22 gates and 2 AO21 gates marked as CG1 to CG5 and two 2-input C-elements marked as C1 and C2), and 4 simple logic gates. Since the AOPT_EO_FA requires less number of gates than the LOPT_EO_FA, the former is area optimized compared to the latter. In Fig. 9, *m*1, *m*2, *int*1, *int*2, *int*3, *nsum*1 and *nsum*0 are the intermediate outputs, with *nsum*1 and *nsum*0 being logically equivalent to SUM1 and SUM0 respectively. Internal nodes *n*1, *n*2 and *n*3 represent isochronic fork junctions. An AO21 gate with inputs P, Q, R, and output S implements the logic function: $S = PQ + R$. Note that the AO21 gate is used to produce the carry output COUT1/COUT0 in the case of LOPT_EO_FA, while in the case of AOPT_EO_FA; the AO22 gate is used to produce the carry output. Since the AO21 gate requires only 8 transistors compared to the requirement of 10 transistors for the AO22 gate for realization in static CMOS style, the LOPT_EO_FA would feature reduced latency for carry output production compared to the AOPT_EO_FA. This is substantiated by the simulation results given in section "Results and discussion". Let us now consider carry-propagate, carry-generate, and carry-kill conditions to explain the operation of LOPT_EO_FA. The LOPT_EO_FA is shown in Fig. 9.

- *Carry- propagate condition*: $A0 = B1 = 1$ (or) $A1 = B0 = 1$
  For either of the specified input combinations, the complex gate CG1 will be enabled and the internal output *int*2 will become 1. The internal completion detector (ICD) will be enabled and *int*3 will also become 1. Assuming CIN1 as 1, the intermediate output *nsum*0 would also become 1. The C-element, marked as C2, would wait for the arrival of rising transitions on *nsum*0 and *int*3, and after their arrival would produce 1 on SUM0. With *int*2 and CIN1 being 1, the carry output COUT1 would become 1. For the inputs combination assumed, both the sum and carry outputs are found to be dependent upon the arrival of all the primary inputs. In a subsequent RTZ phase, even with A0 and/or B1 or A1 and/or B0 returning to 0, COUT1 could assume the spacer state irrespective of CIN1 assuming the spacer. Also, SUM0 could RTZ regardless of CIN1 returning to 0. These demonstrate the early reset nature of LOPT_EO_FA.
- *Carry-generate condition*: $A1 = B1 = 1$
  In this case, the complex gate CG4 will be enabled since A1 and B1 are 1, which will result in the generation of the carry output $COUT1 = 1$ irrespective of whether CIN1 or CIN0 is 1. Further, a rising transition on *m*1 would be followed by a rising transition on *int*1 which would be further followed by a similar rising transition on *int*3. Depending on whether CIN0 or CIN1 is asserted as 1, *nsum*0 or *nsum*1 would experience a respective rising transition, which when coupled with the rising transition on *n*3 results in either SUM0 or SUM1 to be asserted high respectively. In the following RTZ phase, with just A1 and/or B1 returning to 0, both COUT1 and SUM0/SUM1 could RTZ signifying the early reset nature of LOPT_EO_FA.

- *Carry-kill condition*: A0 = B0 = 1

  In this case, the complex gate CG5 will be enabled since A0 and B0 are 1, which will eventually result in COUT0 = 1, regardless of whether CIN1 or CIN0 is 1. Further, the rising transitions on *m*2 and *int*1 would be followed by a similar rising transition on *int*3. Depending on whether CIN0 or CIN1 is asserted as 1, *nsum*0 or *nsum*1 would experience a rising transition, which when coupled with the rising transition on *n*3 results in either SUM0 or SUM1 to be asserted high. In the subsequent RTZ phase, if A0 and/or B0 returns to 0, both COUT0 and SUM0/SUM1 could RTZ which demonstrates the early reset nature of LOPT_EO_FA.

Both the AOPT_EO_FA and LOPT_EO_FA would wait for the arrival of all the valid primary inputs to produce the sum output, while the carry output may be produced in an early output fashion depending upon whether the carry-generate or carry-kill condition is activated. In the RTZ phase, even with any one dual-rail primary input of the full adders assuming the spacer state, all the internal outputs and nodes could RTZ, and subsequently both the sum and carry outputs could be reset irrespective of the other inputs becoming spacers thus highlighting the early output viz. early reset nature of AOPT_EO_FA and LOPT_EO_FA. Hence, when AOPT_EO_FA and LOPT_EO_FA are duplicated and cascaded to form an RCA, application of the relative-timing assumption on the internal carries would become necessary in relation to the primary sum outputs as described in section "Early output logic and relative-timing" (refer Fig. 7).

From the library information (Synopsys 2012), the extent of relative-timing assumption can be theoretically estimated for the AOPT_EO_FA and LOPT_EO_FA based RCAs neglecting any small wire delays. Let us consider only the minimum size gates of the library (Synopsys 2012) for this discussion. If the AOPT_EO_FA is cascaded to form an RCA, during the RTZ phase, the delay involved in the direct reset of the sum output of any full adder stage would be approximately equal to the sum of the propagation delays of two AO22 gates and one 2-input C-element, which equates to 0.250 ns. On the other hand, the delay encountered for the indirect reset of the sum output of any full adder stage through the production and propagation of the carry signal from a previous full adder stage would be approximately equal to the sum of the propagation delays of three AO22 gates and one 2-input C-element, which equals 0.322 ns. Thus a timing slack, i.e., a relative-timing assumption of 0.072 ns would be implicit in the AOPT_EO_FA based RCA. Based on a similar calculation, the timing slack implicit in the LOPT_EO_FA based RCA is found to be 0.025 ns. For the relative-timed RCA constructed using the early output version of Seitz's weak-indication full adder, shown in Fig. 5, the timing slack is estimated to be 0.133 ns, which is 85 and 432 % higher than the respective timing slacks of AOPT_EO_FA and LOPT_EO_FA based relative-timed RCAs. It may be noted that the small relative-timing assumption(s) implicit in the AOPT_EO_FA based RCA and the LOPT_EO_FA based RCA are independent of the RCA size. The timing slacks calculated may be reduced through selective usage of bigger size gates in select areas of the carry output logic of the proposed full adders.

With respect to carry-generate or carry-kill condition alone existing in the AOPT_EO_FA based RCA, the time taken for direct RTZ of the sum output of a full adder stage equals the sum of propagation delays of two AO22 gates and one 2-input C-element. The time taken for indirect RTZ of the sum output of a full adder stage based on the carry input provided by a preceding full adder stage also involves the sum of propagation delays of two AO22 gates and one 2-input C-element. Hence due to delay-balancing of the signal paths, there is no relative-timing assumption required as such for concurrent activation of the full adders in the AOPT_EO_FA based RCA. For non-concurrent activation of the full adders though, a relative-timing assumption equal to the delay of an AO22 gate would be required. This is indeed similar to the relative-timing assumption mandated for the carry-propagate condition as mentioned earlier.

Now presuming that only the carry-generate or carry-kill condition occurs in the LOPT_EO_FA based RCA, the time taken for direct RTZ of the sum output of a full adder stage is equal to the sum of the propagation delays of a 2-input AND gate, a 2-input OR gate, an AO22 gate, and a 2-input C-element. The time taken for indirect RTZ of the sum output of a full adder stage based on the carry input provided by a preceding full adder stage equals the sum of propagation delays of a 2-input AND gate, an AO21 gate, an AO22 gate, and a 2-input C-element. The timing slack calculated equates to just 0.003 ns. On the other hand, the timing slack calculated earlier for the carry-propagate condition is 0.025 ns. Hence, it can be inferred that the timing slack calculation is pertinent to the carry-propagate condition and not to the carry-generate/carry-kill conditions since the timing slack for the latter would be subsumed in the timing slack provision made for the former. In this context, it should be noted that the carry-generate/carry-kill condition does not exist unanimously in an RCA, and typically the carry-propagation occurs between at least 4 full adder stages in an RCA as observed in (Garside 1993).

## Results and discussion

### Simulation results of different asynchronous RCAs: power, latency, and area

Several 32-bit asynchronous RCAs were constructed in a semi-custom design fashion at the gate-level by utilizing different strong-indication, weak-indication, and early output asynchronous full adders separately. The structural integrity of the different asynchronous full adders and RCAs was preserved during physical realization (i.e., technology mapping) to pave the way for legitimate comparison after synthesis. They were implemented using the elements of the 32/28 nm digital cell library (Synopsys 2012). The 2-input C-element was alone designed manually using the AO222 gate by incorporating feedback and was made available to realize the various asynchronous full adders and RCAs, registers, and the completion detector. Logic decomposition of higher fan-in C-element(s), where necessary, was performed according to the safe quasi-delay-insensitive logic decomposition procedure put forward in (Balasubramanian and Mastorakis 2011). The asynchronous system implemented, as shown in Fig. 1, comprises the RCA for the function block, the input registers, and the completion detection circuit. The input registers and the completion detector of various RCAs are identical, and only their asynchronous function blocks (i.e., RCAs) differ in their physical composition. Hence the differences between the simulation results obtained for the various asynchronous RCAs can be directly attributed to the physical differences between their constituent full adders.

More than 1000 random input vectors were applied to the asynchronous RCAs at time intervals of 20 ns through test benches in order to capture their switching activities. The generated .vcd files were subsequently used for average power estimation using Synopsys tool (PrimeTime). Only the worst-case forward latency could be evaluated since the EDA tool basically determines the critical path (i.e., longest carry propagation path) timing. The maximum forward latency signifies the maximum carry propagation delay encountered for the application and processing of valid data, as highlighted by the data path logic in blue line in Fig. 6. For application of spacer data, the carry propagation delay may or may not be significant depending upon whether the RCA is strongly indicating, weakly indicating, early output, or relative-timed, as mentioned in Table 1. Note here that Table 1 has already highlighted the timing efficiency of the relative-timed RCA versus other asynchronous RCAs.

As part of advanced timing analysis, a virtual clock was used just to constrain the input and output ports of the RCAs, and it did not contribute to any power dissipation since it was only virtually made available as part of the designs. Appropriate wire loads (i.e., parasitics) were automatically included whilst performing the simulations to estimate timing and power. The power, latency, RCA area, and the constituent full adder area of different asynchronous RCAs are given in Table 2.

The indication type of each asynchronous RCA is highlighted in the 1st column of Tables 2 and 3. The logic element(s) recurring in the critical carry-propagation path of different RCAs are mentioned in Table 3. Although (Balasubramanian 2011) presented an early output full adder design, when it is cascaded to form a RCA, the RCA type is also early output. In contrast, the proposed AOPT_EO_FA and LOPT_EO_FA though being early output full adders when cascaded leads to relative-timed RCAs. The simulation results in Table 2 correspond to a typical case PVT specification (1.05 V, 25 °C) of the 32/28 nm CMOS process (Synopsys 2012).

It may be understood that the difference in the latency values of the various RCAs, as seen in Table 2, is a direct consequence of the diverse elements recurring in their critical paths as shown in Table 3. It is important to note that less the number of logic element(s) and less complex the logic element(s) are in the critical path would imply reduced latency. Also, the difference between the area metrics of the various RCAs is attributable to the difference between the area metrics of the constituent full adders, with the areas of different full adders given in the 5th column of Table 2. It is clear from Table 2 that the AOPT_EO_FA has the least area occupancy of 23.89 μm$^2$ in comparison with all the other asynchronous full adders, and the LOPT_EO_FA consumes 15 % more Silicon.

**Table 2 Power, latency, and area of various 32-bit asynchronous RCAs constructed using diverse full adders**

| Full adder constituting the RCA; RCA type is given alongside | Power (μW) | Latency (ns) | RCA area (μm$^2$) | Full adder area (μm$^2$) |
|---|---|---|---|---|
| Singh (1981)—strong | 2190 | 14.61 | 2529 | 54.64 |
| Sparsø and Staunstrup (1993)—strong | 2181 | 9.26 | 2504.60 | 53.88 |
| Toms (2006)—strong | 2172 | 9.04 | 2293.14 | 47.27 |
| Sparsø and Staunstrup (1993)—weak | 2177 | 8.24 | 2423.27 | 51.34 |
| Toms and Edwards (2010)—weak | 2192 | 9.66 | 2642.85 | 58.20 |
| Folco et al. (2005)—weak | 2171 | 7.00 | 2016.63 | 38.63 |
| Balasubramanian and Edwards (2008c)—weak | 2174 | 4.43 | 2097.96 | 41.17 |
| Balasubramanian (2015)—weak | 2171 | 3.32 | 2049.16 | 39.65 |
| Balasubramanian (2011)—early output | 2161 | 3.10 | 1658.80 | 27.45 |
| Seitz (1979) Early output version—relative-timed | 2165 | 5.24 | 1870.25 | 34.06 |
| AOPT_EO_FA (proposed)—relative-timed | 2164 | 3.03 | 1544.94 | 23.89 |
| LOPT_EO_FA (proposed)—relative-timed | 2168 | 2.91 | 1658.80 | 27.45 |

It can be seen in Table 2 that the power dissipation does not vary much across the different RCAs although the variations in their areas may be quite significant. This is because input/output mode asynchronous circuits tend to exhibit a unique signal propagation path for each input pattern unlike synchronous designs since they tend to satisfy the monotonic cover constraint (Sparsø and Furber 2001; Balasubramanian and Edwards 2010)—the monotonic cover constraint specifies that only one product term in a logic expression is active at a time. The power consumption of various constituent full adders tends to be more or less equal. Hence, timing and area parameters of the different RCAs deserve more importance than power dissipation for comparison purpose.

Among the different 32-bit RCAs mentioned in Table 2, the LOPT_EO_FA based RCA features the least latency. This is because the LOPT_EO_FA based RCA comprises only a single gate viz. the AO21 gate in the critical carry propagation path. In comparison, the AOPT_EO_FA based RCA despite featuring a single complex gate in the critical carry propagation path, and in addition being relative-timed as well experiences more latency by 4.1 % since its critical path comprises the AO22 gate, which is expensive by 2 more transistors compared to the AO21 gate used in the LOPT_EO_FA based RCA.

Although the RCA constructed using the weak-indication full adder of (Balasubramanian 2015) and the RCA composed using the early output full adder of (Balasubramanian 2011) also comprise only a single logic gate in the critical carry propagation path of each full adder stage, as shown in Table 3, they suffer from increased latency to the tune of 14.1 and 6.5 % compared to the LOPT_EO_FA based RCA, and occupy more area by

**Table 3 Logic element(s) found recurring in the critical path of different asynchronous RCAs**

| Full adder comprising the RCA; RCA type is given alongside | Critical path logical elements |
|---|---|
| Singh (1981)—strong | 2 CE2, 2 OR3 |
| Sparsø and Staunstrup (1993)—strong | CE2, OR4 |
| Toms (2006)—strong | CE2, 2 OR2 |
| Sparsø and Staunstrup (1993)—weak | CE2, OR3 |
| Toms and Edwards (2010)—weak | AND2, CE2, OR3 |
| Folco et al. (2005)—weak | CE2, OR2 |
| Balasubramanian and Edwards (2008c)—weak | AO222 |
| Balasubramanian (2015)—weak | AO21 |
| Balasubramanian (2011)—early output | AO22 |
| Seitz (1979) Early output version—relative-timed | AND3, OR3 |
| AOPT_EO_FA (proposed)—relative-timed | AO22 |
| LOPT_EO_FA (proposed)—relative-timed | AO21 |

CE2: 2-input C-element; AND2: 2-input AND gate

OR2/3/4: 2/3/4-input OR gate; AO222 and AO21 are complex gates

32.6 and 7.4 % in comparison with the AOPT_EO_FA based RCA. Further, compared to the 32-bit asynchronous RCA implemented using the early output version of Seitz's weak-indication full adder, the proposed early output full adders, viz. LOPT_EO_FA and AOPT_EO_FA based RCAs report respective reductions in latency and area by 44.5 and 17.4 % with negligible or almost nil power overheads.

### Theoretical estimation of cycle time of different asynchronous RCAs subject to various carry-propagation lengths

A 32-bit asynchronous ALU implementation was reported in (Garside 1993) and it was found that addition comprises about 80 % of the operations performed by the ALU. It has been found that about 60 % of the random inputs involve carry propagation to less than or equal to 4 stages in a 32-bit adder, and nearly 100 % of the random inputs entail carry propagation to about 8 stages or less. A majority of the address calculations performed by the ALU involve carry propagation to up to 16 stages, and about 45 % of the data processing operations require carry-propagation over the entire adder width, i.e., 30–32 full adder stages.

Given this, it is clear that the cycle time is an important timing parameter in asynchronous designs employing delay-insensitive codes and utilizing 4-phase handshaking since the cycle time represents the time duration for completing a single data transaction, i.e., the processing of a valid data input followed by the RTZ. Notice that the estimation of cycle time using a commercial (i.e., synchronous) EDA tool is normally infeasible since the EDA tool basically estimates the critical path timing. Hence the cycle time parameter is theoretically (i.e., approximately) calculated by neglecting the small interconnect delays between the full adder stages of the RCAs. To compute the cycle times of different 32-bit asynchronous RCAs commensurate with different carry chain lengths of 4, 8, 16 and 32 full adder stages, the worst-case latencies given in Table 2 are averaged, and the latencies of individual full adder blocks are evaluated. These are then scaled up according to the number of carry propagation stages viz. 4, 8, 16 and 32 full adder stages in accordance with the generic cycle time magnitudes given in Table 1. The calculated cycle times are given in Table 4.

**Table 4 Theoretical (i.e., approximate) cycle time estimates of various 32-bit asynchronous RCAs corresponding to different carry chain lengths**

| Full adder constituting the RCA; RCA type is given alongside | Theoretical cycle time (in ns) for RCAs corresponding to various carry-propagation chain lengths | | | |
|---|---|---|---|---|
| | 4 Full adders | 8 Full adders | 16 Full adders | 32 Full adders |
| Singh (1981)—strong | 29.22 | 29.22 | 29.22 | 29.22 |
| Sparsø and Staunstrup (1993)—strong | 18.52 | 18.52 | 18.52 | 18.52 |
| Toms (2006)—strong | 18.08 | 18.08 | 18.08 | 18.08 |
| Sparsø and Staunstrup (1993)—weak | 2.06 | 4.12 | 8.24 | 16.48 |
| Toms and Edwards (2010)—weak | 2.42 | 4.83 | 9.66 | 19.32 |
| Folco et al. (2005)—weak | 1.75 | 3.5 | 7 | 14 |
| Balasubramanian and Edwards (2008c)—weak | 0.83 | 1.38 | 2.49 | 4.71 |
| Balasubramanian (2015)—weak | 0.62 | 1.04 | 1.87 | 3.53 |
| Balasubramanian (2011)—early output | 0.58 | 0.97 | 1.74 | 3.29 |
| Seitz (1979) Early output—relative-timed | 0.82 | 1.47 | 2.78 | 5.40 |
| AOPT_EO_FA (proposed)—relative-timed | 0.47 | 0.85 | 1.61 | 3.12 |
| LOPT_EO_FA (proposed)—relative-timed | 0.45 | 0.82 | 1.55 | 3 |

Strongly indicating RCAs feature the maximum cycle time, while basic weak-indication and distributed/biased weak-indication asynchronous RCAs (Balasubramanian 2015) have cycle times in descending order. This is followed by a reduced cycle time for the early output full adder based RCA of (Balasubramanian 2011). Overall, the proposed early output full adders based relative-timed RCAs exhibit the least cycle time as can be seen in Table 4. However, the cycle times of the proposed LOPT_EO_FA based relative-timed RCA are slightly lesser than the cycle times of the other proposed AOPT_EO_FA based relative-timed RCA for different carry propagation lengths. This is because the LOPT_EO_FA has reduced latency compared to the AOPT_EO_FA, as seen in Table 2.

In comparison with the least of the cycle time values of various strong-indication, weak-indication, and early output asynchronous RCAs as shown in Table 4, the proposed LOPT_EO_FA based relative-timed RCA facilitates minimum reductions in cycle time to the tune of 83.4, 15 and 8.8 % for carry-propagation over the entire RCA width (i.e., 32 full adder stages), and maximum reductions in cycle time by 97.5, 27.4 and 22.4 % respectively for a typical carry-propagation length of 4 full adder stages. For a similar comparison, the proposed LOPT_EO_FA based relative-timed RCA exhibits 45.1 and 44.4 % reductions in cycle time compared to the Seitz's early output full adder based relative-timed RCA. In comparison with the calculated cycle times of the proposed early output full adders based relative-timed RCAs, the necessary relative-timing assumptions of 0.072 and 0.025 ns appear rather insignificant.

## Conclusions

This article has presented two novel latency/area optimized early output asynchronous full adder designs, viz. the LOPT_EO_FA and the AOPT_EO_FA. Both the LOPT_EO_FA and the AOPT_EO_FA are robust designs since they would guarantee gate-orphan freedom. The proposed early output full adders not only have optimized latency/area but also facilitate less cycle time when incorporated into an RCA thus featuring reduced delay and/or area occupancy without sacrificing the power advantage.

When the LOPT_EO_FA or the AOPT_EO_FA is cascaded to form a RCA structure, relative-timing assumptions have to be employed to overcome any potential problem of gate/wire orphan(s). Imposing of relative-timing assumptions tends to incur only a small theoretical timing slack of 0.072 ns for the AOPT_EO_FA based RCA, and a much smaller timing slack of 0.025 ns for the LOPT_EO_FA based RCA. For an asynchronous RCA constructed using the proposed early output full adders, and with the relative-timing assumptions included, the principal advantages imminent are computation with valid inputs is data-dependent, and computation with spacer inputs involves a very fast constant-time reset of just 1 full adder delay, thus resulting in the least cycle time. The LOPT_EO_FA and AOPT_EO_FA based RCAs besides featuring less cycle times also pave the way for less area and power dissipation in comparison with other asynchronous RCAs of similar size, as demonstrated by the simulation results.

Compared to the 32-bit relative-timed RCA realized using the early output version of Seitz's weak-indication full adder, the proposed early output full adders, i.e., LOPT_EO_FA and AOPT_EO_FA based relative-timed RCAs report respective reductions in (forward) latency and area metrics by 44.5 and 17.4 % with negligible or almost nil power overheads. Further, in comparison with the optimized 32-bit RCAs corresponding to strong-indication, weak-indication, and early output asynchronous types, the LOPT_EO_FA based 32-bit relative-timed RCA achieves respective reductions in (forward) latency by 67.8, 12.3 and 6.1 %, while the AOPT_EO_FA based 32-bit relative-timed RCA enables corresponding reductions in area by 32.6, 24.6 and 6.9 %, with practically no power penalty.

Furthermore, the proposed LOPT_EO_FA based relative-timed RCA enables minimum reductions in cycle time by 83.4, 15, and 8.8 % for carry-propagation over the entire RCA width of 32-bits when compared to the least of the cycle time estimates of various strong-indication, weak-indication, and early output asynchronous RCAs of similar size. It has been reported in (Garside 1993) that the mean value of the longest carry-propagation chain is roughly 4 for 32-bits operand addition. Hence, for carry-propagation over a typical length of 4 full adder stages in the RCA, the maximum reductions in cycle time effected by the proposed LOPT_EO_FA based relative-timed RCA in comparison with the least of the cycle times of various strong-indication, weak-indication, and early output asynchronous RCAs of similar size are 97.5, 27.4, and 22.4 % respectively. On average, considering different carry-propagation lengths of 4, 8, 16, and 32 full adder stages, as mentioned in Table 4, the proposed early output full adders based asynchronous RCAs tend to achieve corresponding reductions in cycle time by 91.9, 17.5, and 11.5 % compared to the optimized average cycle times of various strong-indication, weak-indication, and early output asynchronous RCAs. Also, in comparison with Seitz's early output full adder based relative-timed RCA, the proposed LOPT_EO_FA based relative-timed RCA facilitates 44.3 % mean reduction in the cycle time.

**Author details**
[1] School of Computer Engineering, Nanyang Technological University, 50 Nanyang Avenue, Singapore 639798, Singapore. [2] Department of Computer Science, College of Information Science and Engineering, Ritsumeikan University, Kusatsu, Shiga 525-8577, Japan.